\catcode`\@=11					



\font\fiverm=cmr5				
\font\fivemi=cmmi5				
\font\fivesy=cmsy5				
\font\fivebf=cmbx5				

\skewchar\fivemi='177
\skewchar\fivesy='60


\font\sixrm=cmr6				
\font\sixi=cmmi6				
\font\sixsy=cmsy6				
\font\sixbf=cmbx6				

\skewchar\sixi='177
\skewchar\sixsy='60


\font\sevenrm=cmr7				
\font\seveni=cmmi7				
\font\sevensy=cmsy7				
\font\sevenit=cmti7				
\font\sevenbf=cmbx7				

\skewchar\seveni='177
\skewchar\sevensy='60


\font\eightrm=cmr8				
\font\eighti=cmmi8				
\font\eightsy=cmsy8				
\font\eightit=cmti8				
\font\eightbf=cmbx8				

\skewchar\eighti='177
\skewchar\eightsy='60


\font\ninei=cmmi9
\font\ninesy=cmsy9

\skewchar\ninei='177
\skewchar\ninesy='60


\font\tenrm=cmr10				
\font\teni=cmmi10				
\font\tensy=cmsy10				
\font\tenex=cmex10				
\font\tenit=cmti10				
\font\tensl=cmsl10				
\font\tenbf=cmbx10				
\font\tentt=cmtt10				
\font\tenss=cmss10				
\font\tensc=cmcsc10				
\font\tenbi=cmmib10				

\skewchar\teni='177
\skewchar\tenbi='177
\skewchar\tensy='60

\def\tenpoint{\ifmmode\err@badsizechange\else
	\textfont0=\tenrm \scriptfont0=\sevenrm \scriptscriptfont0=\fiverm
	\textfont1=\teni  \scriptfont1=\seveni  \scriptscriptfont1=\fivemi
	\textfont2=\tensy \scriptfont2=\sevensy \scriptscriptfont2=\fivesy
	\textfont3=\tenex \scriptfont3=\tenex   \scriptscriptfont3=\tenex
	\textfont4=\tenit \scriptfont4=\sevenit \scriptscriptfont4=\sevenit
	\textfont5=\tensl
	\textfont6=\tenbf \scriptfont6=\sevenbf \scriptscriptfont6=\fivebf
	\textfont7=\tentt
	\textfont8=\tenbi \scriptfont8=\seveni  \scriptscriptfont8=\fivemi
	\def\rm{\tenrm\fam=0 }%
	\def\it{\tenit\fam=4 }%
	\def\sl{\tensl\fam=5 }%
	\def\bf{\tenbf\fam=6 }%
	\def\tt{\tentt\fam=7 }%
	\def\ss{\tenss}%
	\def\sc{\tensc}%
	\def\bmit{\fam=8 }%
	\rm\setparameters\setbaselines\fi}


\font\twelverm=cmr12				
\font\twelvei=cmmi12				
\font\twelvesy=cmsy10	scaled\magstep1		
\font\twelveex=cmex10	scaled\magstep1		
\font\twelveit=cmti12				
\font\twelvesl=cmsl12				
\font\twelvebf=cmbx12				
\font\twelvett=cmtt12				
\font\twelvess=cmss12				
\font\twelvesc=cmcsc10	scaled\magstep1		
\font\twelvebi=cmmib10	scaled\magstep1		

\skewchar\twelvei='177
\skewchar\twelvebi='177
\skewchar\twelvesy='60

\def\twelvepoint{\ifmmode\err@badsizechange\else
	\textfont0=\twelverm \scriptfont0=\eightrm \scriptscriptfont0=\sixrm
	\textfont1=\twelvei  \scriptfont1=\eighti  \scriptscriptfont1=\sixi
	\textfont2=\twelvesy \scriptfont2=\eightsy \scriptscriptfont2=\sixsy
	\textfont3=\twelveex \scriptfont3=\tenex   \scriptscriptfont3=\tenex
	\textfont4=\twelveit \scriptfont4=\eightit \scriptscriptfont4=\sevenit
	\textfont5=\twelvesl
	\textfont6=\twelvebf \scriptfont6=\eightbf \scriptscriptfont6=\sixbf
	\textfont7=\twelvett
	\textfont8=\twelvebi \scriptfont8=\eighti  \scriptscriptfont8=\sixi
	\def\rm{\twelverm\fam=0 }%
	\def\it{\twelveit\fam=4 }%
	\def\sl{\twelvesl\fam=5 }%
	\def\bf{\twelvebf\fam=6 }%
	\def\tt{\twelvett\fam=7 }%
	\def\ss{\twelvess}%
	\def\sc{\twelvesc}%
	\def\bmit{\fam=8 }%
	\rm\setparameters\setbaselines\fi}


\font\fourteenrm=cmr12	scaled\magstep1		
\font\fourteeni=cmmi12	scaled\magstep1		
\font\fourteensy=cmsy10	scaled\magstep2		
\font\fourteenex=cmex10	scaled\magstep2		
\font\fourteenit=cmti12	scaled\magstep1		
\font\fourteensl=cmsl12	scaled\magstep1		
\font\fourteenbf=cmbx12	scaled\magstep1		
\font\fourteentt=cmtt12	scaled\magstep1		
\font\fourteenss=cmss12	scaled\magstep1		
\font\fourteensc=cmcsc10 scaled\magstep2	
\font\fourteenbi=cmmib10 scaled\magstep2	

\skewchar\fourteeni='177
\skewchar\fourteenbi='177
\skewchar\fourteensy='60

\def\fourteenpoint{\ifmmode\err@badsizechange\else
	\textfont0=\fourteenrm \scriptfont0=\tenrm \scriptscriptfont0=\sevenrm
	\textfont1=\fourteeni  \scriptfont1=\teni  \scriptscriptfont1=\seveni
	\textfont2=\fourteensy \scriptfont2=\tensy \scriptscriptfont2=\sevensy
	\textfont3=\fourteenex \scriptfont3=\tenex \scriptscriptfont3=\tenex
	\textfont4=\fourteenit \scriptfont4=\tenit \scriptscriptfont4=\sevenit
	\textfont5=\fourteensl
	\textfont6=\fourteenbf \scriptfont6=\tenbf \scriptscriptfont6=\sevenbf
	\textfont7=\fourteentt
	\textfont8=\fourteenbi \scriptfont8=\tenbi \scriptscriptfont8=\seveni
	\def\rm{\fourteenrm\fam=0 }%
	\def\it{\fourteenit\fam=4 }%
	\def\sl{\fourteensl\fam=5 }%
	\def\bf{\fourteenbf\fam=6 }%
	\def\tt{\fourteentt\fam=7}%
	\def\ss{\fourteenss}%
	\def\sc{\fourteensc}%
	\def\bmit{\fam=8 }%
	\rm\setparameters\setbaselines\fi}


\font\seventeenrm=cmr10 scaled\magstep3		


\newdimen\rp@
\newcount\@basestretchnum
\newskip\@baseskip
\newskip\headskip
\newskip\footskip


\def\setparameters{\rp@=.1em
	\headskip=24\rp@
	\footskip=\headskip
	\delimitershortfall=5\rp@
	\nulldelimiterspace=1.2\rp@
	\scriptspace=0.5\rp@
	\abovedisplayskip=10\rp@ plus3\rp@ minus5\rp@
	\belowdisplayskip=10\rp@ plus3\rp@ minus5\rp@
	\abovedisplayshortskip=5\rp@ plus2\rp@ minus4\rp@
	\belowdisplayshortskip=10\rp@ plus3\rp@ minus5\rp@
	\normallineskip=\rp@
	\lineskip=\normallineskip
	\normallineskiplimit=0pt
	\lineskiplimit=\normallineskiplimit
	\jot=3\rp@
	\setbox0=\hbox{\the\textfont3 B}\p@renwd=\wd0
	\skip\footins=12\rp@ plus3\rp@ minus3\rp@
	\skip\topins=0pt plus0pt minus0pt}


\def\setbaselines{\maxdepth=4\rp@\baselinestretch=\@basestretchnum}


\def\baselinestretch{\afterassignment\@basestretch\@basestretchnum}
\def\@basestretch{%
	\@baseskip=12\rp@ \divide\@baseskip by1000
	\normalbaselineskip=\@basestretchnum\@baseskip
	\baselineskip=\normalbaselineskip
	\bigskipamount=\the\baselineskip
		plus.25\baselineskip minus.25\baselineskip
	\medskipamount=.5\baselineskip
		plus.125\baselineskip minus.125\baselineskip
	\smallskipamount=.25\baselineskip
		plus.0625\baselineskip minus.0625\baselineskip
	\setbox\strutbox=\hbox{\vrule height.708\baselineskip
		depth.292\baselineskip width0pt }}



\def\makeheadline{\vbox to0pt{\baselinestretch=1000
	\vskip-\headskip \vskip1.5pt
	\line{\vbox to\ht\strutbox{}\the\headline}\vss}\nointerlineskip}

\def\makefootline{\baselineskip=\footskip\line{\the\footline}}

\def\big#1{{\hbox{$\left#1\vbox to8.5\rp@ {}\right.\n@space$}}}
\def\Big#1{{\hbox{$\left#1\vbox to11.5\rp@ {}\right.\n@space$}}}
\def\bigg#1{{\hbox{$\left#1\vbox to14.5\rp@ {}\right.\n@space$}}}
\def\Bigg#1{{\hbox{$\left#1\vbox to17.5\rp@ {}\right.\n@space$}}}


\mathchardef\alpha="710B
\mathchardef\beta="710C
\mathchardef\gamma="710D
\mathchardef\delta="710E
\mathchardef\epsilon="710F
\mathchardef\zeta="7110
\mathchardef\eta="7111
\mathchardef\theta="7112
\mathchardef\iota="7113
\mathchardef\kappa="7114
\mathchardef\lambda="7115
\mathchardef\mu="7116
\mathchardef\nu="7117
\mathchardef\xi="7118
\mathchardef\pi="7119
\mathchardef\rho="711A
\mathchardef\sigma="711B
\mathchardef\tau="711C
\mathchardef\upsilon="711D
\mathchardef\phi="711E
\mathchardef\chi="711F
\mathchardef\psi="7120
\mathchardef\omega="7121
\mathchardef\varepsilon="7122
\mathchardef\vartheta="7123
\mathchardef\varpi="7124
\mathchardef\varrho="7125
\mathchardef\varsigma="7126
\mathchardef\varphi="7127
\mathchardef\imath="717B
\mathchardef\jmath="717C
\mathchardef\ell="7160
\mathchardef\wp="717D
\mathchardef\partial="7140
\mathchardef\flat="715B
\mathchardef\natural="715C
\mathchardef\sharp="715D


\def\err@badsizechange{%
	\immediate\write16{--> Size change not allowed in math mode, ignored}}

\baselinestretch=1000
\tenpoint

\catcode`\@=12					
\catcode`\@=11
\expandafter\ifx\csname @iasmacros\endcsname\relax
	\global\let\@iasmacros=\par
\else	\immediate\write16{}
	\immediate\write16{Warning:}
	\immediate\write16{You have tried to input iasmacros more than once.}
	\immediate\write16{}
	\endinput
\fi
\catcode`\@=12


\def\rmb{\seventeenrm}

\def\singlespace{\baselineskip=\normalbaselineskip}
\def\halfspace{\baselineskip=1.5\normalbaselineskip}
\def\doublespace{\baselineskip=2\normalbaselineskip}


\def\AB{\bigskip\parindent=40pt
        \centerline{\bf ABSTRACT}\medskip\halfspace\narrower}
\def\AE{\bigskip\nonarrower\doublespace}
\def\nonarrower{\advance\leftskip by-\parindent
	\advance\rightskip by-\parindent}


\def\boxit#1{\vbox{\hrule\hbox{\vrule\kern3pt
	\vbox{\kern3pt#1\kern3pt}\kern3pt\vrule}\hrule}}

\def\hence{\leavevmode\hbox{\bf .\raise5.5pt\hbox{.}.} }

\def\dalemb#1#2{{\vbox{\hrule height.#2pt
	\hbox{\vrule width.#2pt height#1pt \kern#1pt \vrule width.#2pt}
	\hrule height.#2pt}}}
\def\gtorder{\mathrel{\raise.3ex\hbox{$>$}\mkern-14mu
             \lower0.6ex\hbox{$\sim$}}}
\def\ltorder{\mathrel{\raise.3ex\hbox{$<$}\mkern-14mu
             \lower0.6ex\hbox{$\sim$}}}

\newdimen\fullhsize
\newbox\leftcolumn
\def\twoup{\hoffset=-.5in \voffset=-.25in
  \hsize=4.75in \fullhsize=10in \vsize=6.9in
  \def\fullline{\hbox to\fullhsize}
  \let\lr=L
  \output={\if L\lr
        \global\setbox\leftcolumn=\columnbox\global\let\lr=R \advancepageno
      \else \doubleformat \global\let\lr=L\fi
    \ifnum\outputpenalty>-20000 \else\dosupereject\fi}
  \def\doubleformat{\shipout\vbox{
    \fullline{\box\leftcolumn\hfil\columnbox}\advancepageno}}
  \def\columnbox{\leftline{\vbox{\makeheadline\pagebody\makefootline}}}
  \tolerance=1000 }
\twelvepoint
\doublespace
{\nopagenumbers{
\rightline{~~~January, 2004}
\bigskip\bigskip
\centerline{\rmb Further thoughts on supersymmetric $E_8$}
\centerline{\rmb as a family and grand unification theory}
\medskip
\centerline{\it Stephen L. Adler
}
\centerline{\bf Institute for Advanced Study}
\centerline{\bf Princeton, NJ 08540}
\medskip
\bigskip\bigskip
\leftline{\it Send correspondence to:}
\medskip
{\singlespace\leftline{Stephen L. Adler}
\leftline{Institute for Advanced Study}
\leftline{Einstein Drive, Princeton, NJ 08540}
\leftline{Phone 609-734-8051; FAX 609-924-8399; email adler@ias.
edu}}
\bigskip\bigskip
}}
\vfill\eject
\pageno=2
\AB
We continue the analysis of the possibility of supersymmetric $E_8$ as 
a family unification and grand unification theory, this time under the 
assumptions that there is a vacuum gluino condensate, but that this  
condensate 
is {\it not} accompanied by dynamical generation of a mass gap in the   
pure $E_8$ gauge theory.    Arguments 
supporting these assumptions are given.  When the $E_8$ theory is coupled 
to supergravity, assuming vanishing of the cosmological constant and a 
supersymmetry breaking scale of around a TeV,  we show that the gluino mass 
induced by gravitational coupling to the condensate is of order $10^{-3}$ eV 
or smaller, compatible with the fermion (and 
particularly the neutrino) mass spectrum.  
We suggest that composite scalar Higgs superfields can arise from a 
chiral glueball in the attractive 3875 channel (and possibly 
other channels), permitting the   
breaking of the original $E_8$ gauge group to a $SO(10)$ grand 
unification group, times a $SU(3)$ family symmetry group and an extra 
$U(1)$ factor.  A general analysis of the Higgs superpotential shows that 
in the absence of gravitational couplings, there is always a supersymmetric 
vacuum in the (unphysical) limit of infinite Higgs superfields.  However, 
when gravitational couplings are included, dimensional analysis of the 
superpotential shows that the vacuum can be stabilized for finite 
Higgs superfields, with the occurrence of dynamical ${\cal F}$-type 
supersymmetry breaking.  
We conclude that an $E_8$ unification  may be theoretically 
viable, providing  
an alternative paradigm for low energy model building, in which the 
supersymmetric partners of the standard model fermions are vectors, and 
in which the only chiral superfields are symmetry breaking Higgs fields.

\AE
\bigskip\bigskip
\vfill\eject
\pageno=3
\centerline{\it 1.~~Introduction}
In a recent publication [1] we gave a mini-review of 
the literature suggesting 
$E_8$ as a grand unification and family unification group, and discussed 
a scenario for the realization of these ideas 
based on the Kovner--Shifman [2]
proposal for a condensate-free vacuum state in supersymmetric Yang--Mills 
theory.  Subsequently, the study of the chiral ring in supersymmetric 
Yang--Mills theory has ruled out the Kovner--Shifman state for the cases 
of the classical non-Abelian groups [3,4] and the 
exceptional group $G_2$ [5],   
and it seems likely that this analysis will eventually extend to the other 
exceptional groups, including $E_8$.  Hence the scenario sketched 
in Ref. [1] is unlikely to work, and we must examine whether a unification 
model based on $E_8$ is compatible with a gluino condensate  
structure [6] for the supersymmetric Yang--Mills vacuum.  This is the issue 
addressed in the present paper.  

We begin with a brief review of the 
advantages, and problems, associated with $E_8$ unification.  We then 
argue that even when a vacuum condensate is present, the conventional 
assumption that there is also a dynamically generated mass gap may fail, 
partly for reasons that are special to the group $E_8$, thus motivating  
our assumption that in the absence of gravitational couplings, 
there is no dynamically 
generated mass gap.  We then 
analyze the implications of the gluino condensate picture when 
the Yang--Mills action is supplemented by the supersymmetric effective 
action arising from the coupling to linearized supergravity.  We next discuss 
the possibility of dynamical generation of a chiral composite superfield, 
and then give a scenario for the dynamical breaking of supersymmetry and 
gauge symmetries, based on an 
augmented superpotential structure for this composite  
permitted in the presence of gravitation.   
We conclude by summarizing the implications of 
our analysis for low energy 
supersymmetric model building based on $E_8$ unification.

\bigskip
\centerline{\it 2.~~Attractive features, and problems, of $E_8$ unification}
\bigskip
In the conventional MSSM approach to supersymmetrization of the standard 
model, the quarks and leptons are placed in a chiral supermultiplet, while 
the gauge bosons are placed in a vector supermultiplet.  Thus, this approach  
does not exploit the possibility afforded by supersymmetry of unifying the 
quarks and gauge bosons of the standard model into a single supermultiplet.  
As sketched in Ref. [1], to which the reader is referred for extensive  
references to the prior literature, a natural candidate theory in which the  
standard model fermions and gauge bosons are superpartners of one another  
is supersymmetric Yang--Mills based on the group $E_8$.  

The attractive features of $E_8$ unification may be briefly summarized 
as follows:
\medskip
\item{1.~} The exceptional group $E_8$ is the unique simple Lie group in 
which the adjoint representation, of dimension 248, is also the fundamental 
representation.  Hence applying the natural grand unification paradigm of  
placing left-handed Weyl spinors in the fundamental, and gauge bosons in 
the adjoint, leads automatically [7] to a  supersymmetric 
Yang--Mills theory in 
which the ``matter'' fermions and the ``gluons'' which bind them are 
superpartners.  This theory is automatically free of gauge anomalies. 
\medskip
\item{2.~}  Under the breaking chain   
$E_8 \supset SU(3) \times E_6$, the 248 of $E_8$ branches [8] as 
$$248=(8,1)+(1,78)+(3,27)+(\overline 3,\overline{27})~~~,\eqno(1a)$$
while under $E_8 \supset SU(3) \times SO(10) \times U(1)$, the 248 
branches as 
$$\eqalign{
248=&(1,16)(3)+ (1,\overline{16})(-3) + (3,16)(-1)
+(\overline 3,\overline{16})(1)+(3,10)(2)\cr
+&(\overline 3,10)(-2)+(3,1)(-4)
+(\overline 3,1)(4)+(8,1)(0)+(1,45)(0)+(1,1)(0)~~~,\cr
}\eqno(1b)$$
with the $U(1)$ generator in parentheses.  Hence, the 248 of $E_8$ naturally 
contains three 27's of $E_6$ and three 16's of $SO(10)$, and so can unify 
the three families [9] into a single representation encompassing a $SO(10)$ 
grand unification group and a $SU(3)$ family group.
\medskip
\item{3.~}  The beta function for supersymmetric Yang--Mills theory with 
an additional multiplet of scalar fields is given by 
$$\beta= {-3 g^3 \over 16 \pi^2} ( C_{\rm adjoint} 
- {1\over 18} C_{\rm scalar})~~~,\eqno(2a)$$
with $C$ one half of the index $\ell$ tabulated in Ref. [8].  Since for 
$E_8$
$${C_{3875} \over C_{248}}= 25 > 18~~~,\eqno(2b)$$
the theory is no longer asymptotically free if any Higgs scalars capable of 
breaking $E_8$ are present [10].  Hence an asymptotically free  
theory is obtained only 
if the gauge symmetry is broken dynamically.  In the absence of gravity,  
the theory is thus a one 
parameter theory, governed by the value of the running coupling at some 
high mass scale below the Planck mass.  

The reason that unification through supersymmetric $E_8$ has been largely 
ignored for the last decade is that there are also three potentially serious 
problems:
\medskip
\item{1.~}Because $248 \times 248$ of $E_8$ contains an $E_8$ singlet, 
a gluino condensate $\langle{\rm Tr} \overline{\chi} \chi \rangle$ can form, 
and is in fact expected under the standard picture [6] of the condensate 
structure of supersymmetric Yang--Mills theories.  If this condensate is 
accompanied by a mass gap of order the $E_8$ scale mass, as is generally  
argued to be the case, then light gluinos cannot appear and 
$E_8$ unification is ruled out. 
\medskip
\item{2.~}Because of well-established ``no-go'' theorems [11] and structural 
restrictions on the induced effective action [12],  
supersymmetric Yang--Mills theory in isolation cannot be 
spontaneously broken.  
\medskip
\item{3.~}Finally, there are well-known phenomenological problems 
with having mirror families, associated with the observed values of the 
electroweak $S$ and $T$ parameters [13].  

In this article we will address the first two problems on this list, by  
arguing that there are scenarios that are capable of surmounting them.   
Hence in principle, supersymmetric $E_8$ 
unification is not ruled out.  This provides a motivation for a future  
detailed analysis, not attempted here,  
to address the third problem listed above.  

\bigskip
\centerline{\it 3.~~Does supersymmetric $E_8$ necessarily have a mass gap?}
\bigskip

Whereas the presence of a gluino condensate in supersymmetric Yang--Mills 
is now reasonably well established, the belief that this condensate 
should be accompanied by a mass gap is less well founded.  One of the 
main arguments for a mass gap is based on the analogy with QCD:  Since 
QCD is a strongly coupled gauge theory with both a chiral condensate 
and a mass gap, one would expect that supersymmetric Yang--Mills theories, 
which are also strongly coupled gauge theories, should have analogous 
features.  However, a number of criticisms can be leveled at this 
analogy, particularly when applied to the $E_8$ gauge group.  

{}First of all, in QCD the mass gap does not take the form of 
Lagrangian mass terms (so called ``current'' mass terms) for the quarks 
of the form $m_q \overline q q$, but rather appears through effective 
masses (so called ``constituent'' masses) when the quarks are bound in 
hadrons.  To see that no ``current'' masses are generated, we note that the  
presence of a chiral condensate in QCD is an indication that there is an 
effective potential for $\overline q q$ with a minimum at the vacuum 
expectation $\langle \overline q q \rangle$.  Since this potential is 
expected to be quadratic around the minimum, substituting 
$\overline q q = N[\overline q q] +  \langle \overline q q \rangle$into  
the effective potential, 
with $N[~~]$ denoting the normal ordered product, one finds an operator 
correction near the minimum proportional to $N[\overline q q]^2$, which 
is not a mass term but rather an effective four fermion interaction.  Thus 
expansion around the minimum associated with the chiral condensate does 
not produce a Lagrangian mass term.  This argument carries over directly 
to the Veneziano--Yankielowicz [6] effective potential for supersymmetric 
Yang--Mills theory,  and shows that the presence of a gluino condensate  
does not imply the generation of Lagrangian mass terms for the gluinos.   

This still leaves the possibility that supersymmetric Yang--Mills theory 
develops gluino effective masses analogous to the ``constituent'' masses 
in QCD, which are generated as a result of the confinement of quarks 
in hadrons. However,  supersymmetric Yang--Mills is 
more closely analogous to QCD with adjoint quarks, than it is to standard 
QCD with fundamental quarks.  In QCD with adjoint quarks the string 
tension vanishes, because the adjoint quark charges can be screened 
by adjoint gluons [14]; in other words, QCD with adjoint quarks is in a Higgs 
phase rather than a confining phase.  For gauge groups other than $E_8$, 
a QCD analog can be created from supersymmetric Yang--Mills 
by adding a chiral multiplet in the fundamental
representation, giving a probe to study the pure Yang--Mills limit.  
This method 
does not give a QCD analog in the case of $E_8$, because the fundamental 
representation is identical to the adjoint representation, and so chiral 
fundamental ``matter'' is also screened rather than confined.  
The same is true for probes using chiral ``matter'' in higher $E_8$ 
representations, since these representations are contained in multiple 
tensor products of the fundamental, and so their charges can be screened  
by the accumulation of multiple $E_8$ gluons.  
(These remarks are in accord with the study by Acharya [15] of confinement 
in supersymmetric Yang--Mills using $M$--theory methods, since in the $E_8$ 
case his method agrees with the trivial, nonconfining center of the 
$E_8$ group.) To sum up, $E_8$ 
supersymmetric Yang--Mills is not a confining theory, and thus it is 
not at all evident that the QCD analogy can 
be used to infer the generation of gluino constituent masses.  

Other arguments for the presence of a mass gap in supersymmetric Yang--Mills 
theory are based on extrapolations from theories of different types,  
such as $N=2$ supersymmetric Yang--Mills theory in four dimensions, or 
string theories or $M$--theory in higher dimensions. An example of the latter  
is the paper of Atiyah and Witten [16], which presents arguments for a 
continuous $M$--theory curve connecting a region of parameter space 
with a mass gap to a limiting region  
related to  four-dimensional supersymmetric Yang--Mills theory.  
Here the problem is that there is 
no guarantee that the mass gap does not vanish in the course 
of the limiting operation needed to recover $N=1$ supersymmetric Yang--Mills 
in four dimensions from a larger, qualitatively different theory.  This could 
happen if either the limit of Yang--Mills theory is on a phase boundary, or 
even if no change of phase is involved, if the mass gap vanishes while the 
phase-defining order parameter (such as the gluino condensate) does not.  
In terms of the superconductive analog for condensate formation, the latter 
is just what happens in gapless superconductors [17], where as the impurity 
concentration is increased, the energy gap vanishes over an open interval 
in which the order parameter (the condensate wave function) is non-zero.  

To summarize, not only is there no proof of the existence of a mass gap in 
supersymmetric $E_8$ Yang--Mills (even in the heuristic sense in which 
there is a ``proof'' that there is a gluino condensate), but the arguments 
advanced for existence of a mass gap are on somewhat shaky ground.  Hence 
we shall make the assumption, in the remainder of this paper, that  
$E_8$ supersymmetric Yang--Mills theory is an exception to the conventional 
lore, and has no intrinsic mass gap.  

\bigskip
\centerline{\it 4.~~The gluino condensate and gluino mass 
in the presence of supergravity}
\bigskip
If nature is supersymmetric, one expects a supersymmetric unified matter 
dynamics to be coupled to supergravity.  To leading order in the 
gravitational coupling $\kappa^2=8\pi G_{\rm Newton}
=M_{\rm Planck}^{-2}$, the effects 
of supergravity on the matter sector can be summarized by an effective 
action $S_{\rm eff~grav}$, given [18] by 
$$\eqalign{
S_{\rm eff}=&\kappa^2 \int d^4x \left[-{3\over 16} j_{\mu}^{(5)}j^{\mu(5)}
+{1\over 48} (P^2+Q^2)\right]\cr
+&\kappa^2  \int d^4x d^4y \left[ {1\over 4}\theta^{\nu\tau}(x) 
(\eta_{\nu\alpha}\eta_{\tau\beta}+ \eta_{\nu\beta}\eta_{\tau\alpha}
-\eta_{\nu\tau}\eta_{\alpha\beta})\Delta_F(x-y) 
\theta^{\alpha\beta}(y)\right.\cr
-&\left. {1\over 8}  \overline{j}_{\tau}(x) 
\left(\eta^{\tau\nu} \gamma \cdot \partial_{x}+{1\over 2} \gamma^{\tau} 
\gamma \cdot \partial_{x} \gamma^{\nu}\right) \Delta_F(x-y) j_{\nu}(y)
\right]~~~.\cr
}\eqno(3a)$$
Here $\Delta_F$ is the massless Feynman propagator, while  
$P,Q,j_{\mu}^5,j_{\nu},\theta_{\mu\nu}$ are the components of the 
matter 
supercurrent, and it is straightforward to verify that Eq.~(3a) is  
supersymmetry invariant when the conservation relations 
$$\partial_{\mu}\theta^{\mu\nu}=\partial_{\nu}\theta^{\mu\nu}
=\partial_{\mu}j^{\mu}=0~~~\eqno(3b)$$ 
are used.  The total action will then be given by 
$$S=S_{\rm matter} + S_{\rm eff~grav}~~~,\eqno(4)$$
with $S_{\rm matter}$ the supersymmetric matter action.  For  
the moment we leave the form of $S_{\rm matter}$ unspecified, until we make 
statements further on 
that specifically assume a supersymmetric Yang--Mills form.   

Let us now consider the vacuum energy implied by Eqs.~(3a) and (4).  
Lorentz invariance implies that the only nonvanishing vacuum expectations  
of components of the current supermultiplet are 
$\langle P \rangle$, $\langle Q \rangle$ and $\langle \theta_{\mu\nu} \rangle
=\langle \theta_0^0 \rangle \eta_{\mu\nu}$, with 
$\eta_{\mu\nu}$ the Minkowski metric. Since we expect supersymmetry to 
be broken in the matter sector, the 
positive semidefinite matter 
vacuum energy density $\langle \theta_0^0 \rangle$ will be nonzero.  Adding  
$\langle \theta_0^0 \rangle$ to the gravitational contributions coming  
from Eq.~(3a), the total vacuum energy 
density becomes  
$$\rho_{\rm VAC}= \langle \theta_0^0\rangle 
-{\kappa^2 \over 48} (\langle P \rangle^2 + \langle Q \rangle^2)~~~.
\eqno(5)$$
By a current algebra calculation using the supersymmetry algebra, one 
can show [19] that the second order gravitino self-energy induced by the 
expectations $\langle P \rangle,\langle Q \rangle$  takes the form 
$$\eqalign{
\Delta S_{\rm mass}=&
{1\over 2} m
\int d^4x \overline{\psi}_{\mu}  (x) \sigma^{\mu\rho}  \psi_{\rho}(x)
+{1\over 2} m^{\prime}
\int d^4x \overline{\psi}_{\mu}  (x)i 
\gamma_5 \sigma^{\mu\rho} \psi_{\rho}(x)
~~~,\cr
m=&{\kappa^2 \over 12} \langle P \rangle~,~~~~
m^{\prime}={\kappa^2 \over 12} \langle Q \rangle~~~,\cr
}\eqno(6)$$
with $\psi_{\mu}$ the gravitino field.  
The condition for the vacuum energy density or cosmological constant
$\rho_{\rm VAC}$ of Eq.~(6) to vanish by cancellation 
between the matter and supergravity contributions is then [19] 
$$\kappa \left[ \langle \theta_0^0 \rangle \over 3 \right]^{1\over 2} = 
{\kappa^2 \over 12}(\langle P \rangle^2 + \langle Q \rangle^2)^{1\over 2}=
(m^2 + m^{\prime~2})^{1\over 2}~~~.\eqno(7a)$$
When the CP nonconserving expectation  
$\langle Q \rangle \propto m^{\prime}$ is zero, this reduces to the Deser- 
Zumino formula [20] for the gravitino mass $m$, 
$$ m={\kappa^2 \over 12} \langle P \rangle=
\kappa \left[ \langle \theta_0^0 \rangle \over 3 \right]^{1\over 2}~~~.
\eqno(7b)$$
If we assume that the matter supersymmetry breaking scale is of order 
a TeV  (and thus, we assume no hidden sectors with higher 
supersymmetry breaking scales), Eq.~(7b) leads to the estimate  
$$\langle P \rangle^{1\over 3} \sim  10^9 {\rm GeV}~~~~\eqno(8a)$$ 
for the 
mass scale associated with $\langle P \rangle$, and to the estimate
$$m \sim 10^{-13}{\rm GeV} =10^{-4}{\rm eV}~~~\eqno(8b)$$ 
for the gravitino mass.  This latter    
is compatible with the current accelerator bound [21] 
of $m \geq 3 \times 10^{-13} 
{\rm GeV}$ for the gravitino mass.  In other words, in great generality, 
if one assumes cancellation of the cosmological constant between the 
matter and supergravity sectors, and a matter supersymmetry breaking 
scale of a TeV, one concludes that the gravitino must be superlight.  

Let us now specialize to the case in which $S_{\rm matter}$ is the 
supersymmetric Yang--Mills action, and estimate the corresponding gluino 
condensate and gluino mass.  For supersymmetric Yang--Mills, the tree level 
operator $P$ is zero, with a contribution first appearing from the 
anomaly supermultiplet given by 
$$P={\beta(g) \over g} {\rm Tr} \overline{\chi} \chi~~~,\eqno(9)$$ 
with $\chi$ the gluino field, with $\beta$ the beta function, and with   
${\rm Tr}\overline{\chi}\chi=\sum_A \overline{\chi^A} \chi^A$.  
Thus the estimate of Eq.~(8a) shows that the gluino condensate must be 
nonzero, and gives an estimate of its magnitude.  To get a corresponding 
estimate of the gluino mass induced by 
gravitational couplings to this condensate, we substitute 
$P= \langle P \rangle + {\beta(g) \over g} {\rm Tr} N[\overline{\chi} \chi]$, 
with $N[~~]$ denoting normal ordering, into Eq.~(3a) for the supergravity  
induced effective action.  Linearizing in the normal ordered terms, this 
gives a gluino mass Lagrangian density term of 
$$ {\kappa^2 \over 12} \langle P \rangle {\beta(g) \over g}   
{1\over 2} {\rm Tr} N[\overline{\chi} \chi]~~~,\eqno(10a)$$
corresponding to a gluino mass of 
$$m_{\rm gluino} =  {\kappa^2 \over 12} \langle P \rangle
\left|{\beta(g) \over g}\right|   
= \left|{\beta(g) \over g}\right| m~~~,\eqno(10b)$$ 
a formula familiar [22] from the theory of anomaly mediated supersymmetry 
breaking.  If we assume that $|{\beta(g) \over g}| \leq 10$, then 
Eqs.~(8b) and (10b) give an estimate for the $E_8$ singlet 
mass term,
$$m_{\rm gluino} \sim 10^{-3} {\rm eV}~~~,\eqno(10c)$$
which sets a lower bound for the observed fermion masses. (Other, non-singlet  
dynamically generated mass terms will of course be needed to give a realistic 
fermion mass spectrum.)  This bound is 
compatible with our knowledge of the neutrino mass spectrum, and so we 
conclude that the $E_8$ supersymmetric Yang--Mills vacuum, 
with its gluino condensate but assuming no intrinsic mass gap, is 
compatible with the idea that the fermions of 
the standard model may be the gluinos of a supersymmetric Yang--Mills theory.   
\vfill\eject 
\bigskip
\centerline{\it 5.~~A scenario for dynamical generation of chiral Higgs 
superfields}
                   
\bigskip
If an $E_8$ theory is to describe observed standard model physics, 
three kinds of symmetries must be broken:  (i) gauge symmetry, (ii) 
the discrete symmetries P, C, and CP, and (iii) supersymmetry.  In order 
for these symmetries to be broken by a supersymmetric analog of the Higgs 
mechanism, we need to have both chiral Higgs superfields, and an effective 
potential for these superfields that breaks the symmetries (i)--(iii).  
In this section we address the issue of obtaining the needed superfields, 
and in the next section we shall analyze whether suitable effective 
potentials can be generated.  

Since we have seen that an asymptotically free $E_8$ supersymmetric 
theory cannot have fundamental Higgs bosons, the Higgs fields must occur 
as dynamically generated composites. Letting $\lambda^A$, $A=1,...,248$ be   
the generator matrices for $E_8$, the strength of the vector exchange 
force between two 248 supermultiplets 1 and 2 in a channel with 
group representation $T$ is proportional to 
$$\eqalign{
&\sum_A \lambda_1^A \lambda_2^A = \lambda_1 \cdot \lambda_2  \cr
=&{1\over 2} [(\lambda_1 + \lambda_2)^2 - \lambda_1^2 -\lambda_2^2]  \cr 
= & C_2(T) - 2 C_2(248) ~~~,\cr
} \eqno(11a)$$
with $C_2(T)$ and $C_2(248)$ respectively one half of the Casimirs for 
the representations $T$ and 248.   The representations $T$ that are  
potentially of interest for the formation of dynamical scalar composites are 
the the 1, 3875, 27000, and 30380, since these are the representations 
that appear symmetrically in the decomposition 
$$248\times 248=1_s+ 248_a+3875_s+27000_s+30380_s~~~.\eqno(11b)$$
(Only the symmetrical terms are of interest because 
forming a Lorentz scalar from two anticommuting spinors requires an 
antisymmetric $\epsilon$ factor in the spinor indices, 
as in Eq.~(13b) below, requiring the 
group structure factor to be symmetric [23].)
Using the formula [8] 
$$C_2(T)={N({\rm adjoint}) \over N(T) } {1\over 2} \ell(T) 
={N(248)\over N(T)} C_T~~~,\eqno(11c)$$
with $N(T)$ the dimension of the representation $T$, and $C_T$ as before  
one half of the index $\ell(T)$, 
we find the values 
$$\eqalign{
C_2(1)=&0~,~~C_2(248)=30~,~~C_2(3875)=48~\cr
C_2(27000)=&62~,~~ C_2(30380)=60~~~,
}\eqno(12a)$$
from which we find   
$$\eqalign{
C_2(1)-2C_2(248)=&-60~~~,\cr
C_2(3875)-2C_2(248)=&-12~~~,\cr
C_2(27000)-2C_2(248)=&2~~~,\cr
C_2(30380)-2C_2(248)=&0~~~.\cr
}\eqno(12b)$$
The most attractive channel is the singlet, which as we have already 
seen is expected to contain a glueball condensate.  According to the 
often used most attractive channel (MAC) rule [24], only the most attractive 
channel is supposed to contain a dynamical composite, but we see no  
compelling justification 
for excluding the possibility that other attractive channels 
may have composites as well.  Thus, the 3875 channel, which is also 
attractive, may contain a dynamical composite.  We note furthermore that 
since the 30380 is on the borderline between attractive and repulsive, and 
the 27000 is only weakly repulsive, renormalization effects may make 
one or both of these effectively attractive, leading to the more speculative  
possibility of further composites beyond the 3875.  

Chiral superfields corresponding to the various possible composite channels 
can be written in terms of the chiral gaugino/gluino superfield 
$W_{\alpha}={1\over 2} \sum_A \lambda^A W_{\alpha}^A$.  The usual 
glueball, corresponding to the channel $T=1$, is 
given (with the spinor indices $\alpha,\beta$ summed over) by 
$$\Phi(1,0)\equiv {\rm Tr} W_{\alpha} W_{\beta} \epsilon^{\alpha \beta}  
= \sum_A  W_{\alpha}^A W_{\beta}^A \epsilon^{\alpha \beta} ~~~,
\eqno(13a)$$
and is an $E_8$ gauge invariant chiral superfield.  
Using the $E_8$ Clebsch $(T m|248 A,248 B)$ one can form further chiral    
superfields 
$$\Phi(T,m)\equiv\sum_{A,B}  (T m|248 A,248 B)  
W_{\alpha}^A W_{\beta}^B \epsilon^{\alpha \beta} ~~~,\eqno(13b)$$ 
which are Lorentz scalar chiral superfields transforming as the $m$th 
basis element of the representation $T$.  Hence the formation of non-singlet  
composites can preserve supersymmetry, through the formation 
of dynamical chiral superfields in attractive channels.  
Thus an $E_8$ gauge theory can, in principle, generate chiral Higgs fields 
that can lead to dynamical symmetry breaking when these chiral superfields    
develop non-vanishing vacuum expectation values.

\bigskip
\centerline{\it 6.~~Structure of effective potentials for the  chiral Higgs 
superfields}
\bigskip 
We must now check whether the theory is allowed to dynamically generate 
an effective superpotential function $f$ of the superfields 
of Eqs.~(13a,b) that 
obeys the following requirements:  First, it should be of dimension 3, 
so that its ${\cal F}$ projection is of dimension 4 as required for an 
action density.  Second, under the classical chiral $U(1)$ transformation 
(the $R$-symmetry transformation) which scales $W_{\alpha}$  as 
$W_{\alpha} \to \exp(i\phi) W_{\alpha}$, the terms in the superpotential 
that do not solely involve $\Phi(1,0)$ should scale as $\exp(2i\phi)$. 
In other words, these terms should have an $R$ quantum number 2, so 
that their   contribution to the effective action is independent of $\phi$.  
(The part of the 
superpotential that is solely a function of the singlet glueball 
$\Phi(1,0)$ is affected by the chiral anomaly, leading to an extra logarithm 
in its effective superpotential, as shown by Veneziano and Yankielowicz [6], 
and so do not show simple $R=2$ scaling.)  Finally, as the gauge coupling 
$g$ approaches zero, or equivalently, as the $E_8$ subtraction-independent 
scale mass $M$ given by 
$$M=\mu \exp\left(-{8 \pi^2 \over 3 C_{\rm adjoint}}
{1 \over g^2(\mu)}\right)~~~
\eqno(14)$$ 
approaches zero, the effective superpotential for fixed chiral superfield 
arguments should approach zero.  We shall require this approach to zero 
to be uniform in the nonsinglet chiral superfield arguments $\Phi(T,m)$  
of the effective potential, 
permitting interchange of taking the zero coupling limit with taking 
derivatives acting on these arguments.    
                                                                  
Abbreviating the singlet glueball by $Z=\Phi(1,0)$, with    
corresponding vacuum expectation $\langle Z \rangle$, and abbreviating the 
general non-singlet glueball by $Y_i=\Phi(T,m)$, 
in the absence 
of a coupling to gravitation the most general holomorphic 
superpotential satisfying the first two requirements, and incorporating  
the Veneziano--Yankielowicz superpotential, is 
$$f=Z \log \left( {Z \over e \langle Z \rangle }\right)   
+ Z F[\{ {Y_i \over Z}\}]  ~~~, \eqno(15a)$$
with $F$ a general function of its arguments.  Since the scale of 
$Z$ is set by $\langle Z \rangle \propto M^3$, the third requirement 
will be satisfied if 
$$ Z F[\{ {Y_i \over Z}\}] \to 0 ~~~\eqno(15b)$$ 
uniformly in the nonsinglet chiral superfields $Y_i$ as  
$Z \to 0$.  Differentiating with respect to $Y_i$, this implies 
that 
$${\partial f \over \partial Y_i} = Z {\partial F \over \partial Y_i}~~~
\eqno(15c)$$ 
also approaches zero as $Z \to 0$.  However, since the right hand side 
of Eq.~(15c) is a function of the ratios $\kappa_i=Y_i /Z$, this 
also implies 
that $\partial f/ Y_i \to 0$ as all of the Higgs superfields 
$Y_i$ are uniformly scaled to infinity, or equivalently, as all of the   
$\kappa_i$ are uniformly scaled to infinity.  Since 
$${\partial f \over Z}=  \log \left( {Z \over  \langle Z \rangle }\right)  
 + F[\{\kappa_i\}] -\sum_j \kappa_j {\partial  F[\{\kappa_i\}]
 \over \partial \kappa_j}~~~,\eqno(16a)$$
the equation $\partial f / Z=0$ always has a solution for any 
set  $\{\kappa_i\}$.  At this solution, the potential 
$$V=\left|{\partial f \over Z}\right|^2 
+ \sum_i \left|{\partial f \over \partial Y_i}\right|^2 ~~~\eqno(16b)$$  
vanishes in the limit as all of the $Y_i$ or $\kappa_i$ are scaled to 
infinity, and so there is at least one supersymmetric (although not 
physically realistic) ground state.  
(If the potential $V$ vanishes monotonically as 
the $\kappa_i$ are scaled to infinity, then 
there cannot even be a metastable supersymmetry breaking vacuum for   
finite values of the Higgs superfields).  Therefore, in the absence 
of coupling to supergravity, the $E_8$ gauge theory, even with the 
dynamical generation of Higgs superfields, does not break supersymmetry.  
This conclusion agrees with general expectations, based for example on 
the Witten index [11], that pure supersymmetric Yang--Mills theory cannot 
break supersymmetry.  

The situation changes qualitatively when supergravity couplings are 
included, since then the superpotential $f$ can depend on the 
dimensionless quantity $\kappa M = M/M_{\rm Planck}$, which vanishes
as either the gauge coupling or the gravitational coupling 
approaches zero.  One can then form O'Raifeartaigh [25] 
or ${\cal F}$-type superpotentials that satisfy all of the requirements, 
and break supersymmetry, with a vacuum stabilized at finite values of   
the Higgs superfields.  For example, letting $X$, $Y_1$ and $Y_2$  
be any of the nonsinglet Higgs   
superfields, and letting $\psi(u)$ be any dimensionless 
positive function that vanishes as 
$u \to \infty$, consider 
$$f=Z \log \left( {Z \over e \langle Z \rangle }\right)   
+Y_1  \left[ \psi({X\over Z}) - \kappa^2 M^2\right] 
+Y_2 \psi({X\over Z})  ~~~,\eqno(17a)$$
which satisfies all of our requirements.  
Then we see that $\partial f/\partial X$ vanishes on the surface  
$Y_1+Y_2=0$, and $\partial f/\partial Z$ then vanishes for 
$Z=\langle Z \rangle$, but 
$$ \left| {\partial f \over Y_1} \right|^2 + 
 \left| {\partial f \over Y_2} \right|^2
 =\left| \psi({X\over Z}) - \kappa^2 M^2 \right|^2
 +\left|  \psi({X\over Z})\right|^2 \geq {1\over 2} \kappa^4  M^4~~~, 
 \eqno(17b)$$ 
and so supersymmetry is necessarily broken.  Since the dimension one 
Higgs superfields are obtained by dividing the dimension three 
fields $Y_i$ by the scale mass squared $M^2$, the correctly 
normalized potential is obtained by multiplying Eq.~(17b) by a factor 
of $M^4$, and so equating the supersymmetry breaking scale to a TeV 
gives the estimate $M^8/M_{\rm Planck}^4 \sim (1 {\rm TeV})^4$, which 
gives $M \sim 10^{11}${\rm GeV} for the scale mass.   

The variant of the superpotential of Eq.~(17a) in which the $Y_2$ term 
is omitted preserves supersymmetry, 
but for suitable choices of $X$, $Y_1$ and $\psi$ can spontaneously 
break the $E_8$ gauge symmetry or various discrete symmetries.  Thus, 
supergravity couplings can generate effective superpotentials which 
stabilize the vacuum and break all of the requisite symmetries, with 
the natural appearance of a hierarchy when $\kappa^2M^2$ is small.  
{}Finally, 
we remark that since the kinetic energy comes from $D$ terms in the 
effective action, which are not required to be holomorphic, the 
considerations of $R$ invariance used above place no restrictions, since 
one can always generate $R=0$ terms as the squared modulus of terms 
with nonzero $R$.  Hence even in the absence of gravitational couplings 
there will be effective kinetic terms for the composite Higgs superfields.  

\vfill\eject
\bigskip
\centerline{\it 7.~~Implications for low energy 
supersymmetric model building}
\bigskip
The proposals just sketched for a unification theory based on $E_8$ involve 
many non-perturbative phenomena, and so their verification or falsification 
will be challenging.    However, they suggest an alternative 
paradigm for low energy supersymmetric model building in which the 
supersymmetric partners of the observed fermions are vectors, rather than 
scalars, and in which the only chiral superfields are the Higgs fields needed 
for gauge symmetry breaking.  One test of the proposals we have made is 
to see whether phenomenologically acceptable supersymmetric extensions 
of the standard model can be constructed using this alternative paradigm. 
Such model building should be feasible within a conventional 
perturbative framework, and will be needed to see whether the third  
objection to $E_8$ unification, involving the presence of mirror fermions 
and their contributions to the electroweak precision parameters, can be 
satisfactorily dealt with.  Within the decade, experiments may also  
make decisive statements about both the presence of additional families 
of mirror fermions, and the Lorentz structure (scalar versus vector) 
of squarks and sleptons.  

\bigskip
\centerline{\bf Acknowledgments}
\bigskip

This work was supported in part by the Department of Energy under
Grant \#DE--FG02--90ER40542.  I wish to thank Bobby Acharya, 
Hooman Davoudiasl, 
Ryuichiro Kitano, Graham Kribs, Tianjun Li, 
Hitoshi Murayama, Marc Thormeier, 
 and especially Nathan Seiberg and Edward Witten,  
for helpful conversations or email correspondence.    
This work was also stimulated by remarks made to me 
long ago by Freeman Dyson, Murray Gell-Mann, and Richard Slansky.  
\vfill\eject
\centerline{\bf References}
\bigskip
\noindent
[1]  S. L. Adler, Phys. Lett. B 533 (2002) 121. \hfill \break
\noindent
[2]  A. Kovner and M. Shifman, Phys. Rev. D 56 (1997) 2396. \hfill\break
\noindent
[3]  F. Cachazo, M. R. Douglas, N. Seiberg, and E. Witten, hep-th/0211170.
\hfill\break
\noindent
[4]  E. Witten, hep-th/0302194.  \hfill \break
\noindent
[5] P. Etingof, V. Kac, math.QA/0305175.  \hfill\break
\noindent
[6] H. P. Nilles, Phys. Lett. B 112 (1982) 455; G. Veneziano,   
S. Yankielowicz, Phys. Lett. B 113 (1982) 231. \hfill\break
\noindent
[7] N. S. Baaklini, Phys. Lett. B 91 (1980) 376; I. Bars,  M. G\"unaydin, 
Phys. Rev. Lett. 45 (1980) 859; S. E. Konshtein, E. S. Fradkin, 
Pis'ma Zh. Eksp. Teor. Fiz. 32 (1980) 575, [English translation: JETP Lett. 
32 (1981) 557].  See also M. Koca, Phys. Lett. B 107 (1981) 73, and 
R. Slansky, Ref. [8].  \hfill\break
\noindent
[8] R. Slansky, Phys. Rep. 79 (1981) 1. \hfill\break 
\noindent
[9]  I. Bars, M. G\"unaydin, Ref. [7] ; R. Slansky, Ref. [8]; 
S. M. Barr, Phys. Rev. D 37 (1988) 204.  For a similar observation in the 
context of supersymmetric nonlinear sigma models, see C. L. Ong, Phys. 
Rev. D 31 (1985) 3271; W. Buchm\"uller, O. Napoli, Phys. Lett. B 163 
(1985) 161; K. Itoh, T. Kugo, H. Kunitomo, Prog. Theor. Phys. 75 (1986) 386; 
U. Ellwanger, Nucl. Phys. B 356 (1991) 46.  \hfill\break
\noindent
[10]  I. Bars, M. G\"unaydin, Ref. [7]. \hfill\break   
\noindent
[11]  E. Witten, Nucl. Phys. B202 (1982) 253. \hfill\break
\noindent
[12]  M. Dine, hep-ph/9612389, published in:  Fields, Strings, and 
Duality: TASI 96,  C. Efthimiou and B Greene (Eds.), World Scientific, 
Singapore, 1997; S. Weinberg, Phys. Rev. Lett. 80 (1998) 3702; I. Affleck,    
M. Dine, and N. Seiberg, Nucl. Phys. B241 (1984) 493.  
\hfill \break
\noindent
[13]  M. E. Peskin, T. Takeuchi, Phys. Rev. D46 (1992) 381; J. Erler, 
P. Langacker, in Particle Data Group, Review of Particle Physics,  
available online at http://pdg.lbl.gov/. \hfill\break
\noindent
[14]  See, e.g., S. L. Adler and H. Neuberger, Phys. Rev. D27 (1983) 1960. 
\hfill\break
\noindent
[15]  B. S. Acharya, hep-th/0101206 and hep-th/0011089.  
\hfill\break
\noindent
[16] M. Atiyah and E. Witten, Adv. Theor. Math. Phys. 6 (2003) 1.
\hfill\break
\noindent
[17] See, e.g., A. A. Abrikosov, Fundamentals of the Theory of Metals, 
North Holland, Amsterdam, 1988, pp. 514-515.\hfill\break
\noindent
[18]  S. L. Adler, Ann. Phys. 290 (2001) 11.  The superconformal case, 
with vanishing anomaly supermultiplet, was given earlier in  S. Deser, 
A. Waldron, hep-th/0012014, Sec. IV.  \hfill\break  
\noindent
[19] S. L. Adler, Gen. Rel. Grav. 33 (2001) 1. \hfill\break
\noindent
[20]  S. Deser, B. Zumino, Phys. Rev. Lett. 38 (1977) 1433; see also   
S. Weinberg, The Quantum Theory of Fields, Vol. III, Cambridge University 
Press, Cambridge, 2000, pp. 336-337.  \hfill\break
\noindent
[21]  D. A. Dicus, S. Nandi, Phys. Rev. D  56 (1997) 4166.  \hfill\break
\noindent
[22]  See e.g. S. Weinberg, Ref. [20], p. 339.  \hfill\break
\noindent
[23] See, e.g. M. E. Peskin, in Recent Advances in Field Theory and 
Statistical Mechanics, Les Houches 1982, J.-B. Zuber, R. Stora (Eds.), 
North Holland, Amsterdam, 1984.  \hfill\break
\noindent
[24]  J. M. Cornwall, Phys. Rev. D10 (1974) 500; S. Raby, S. Dimopoulos, 
L. Susskind, Nucl. Phys. B169 (1980) 373.  \hfill\break
\noindent
[25]  L. O'Raifeartaigh, Nucl. Phys. B 96 (1975) 331; see also   
 S. Weinberg, Ref. [20], pp. 83-85. \hfill\break
\bye